\documentclass[fleqn,10pt]{wlscirep}
\usepackage[utf8]{inputenc}
\usepackage[T1]{fontenc}

\DeclareUnicodeCharacter{2500}{?}

\newcommand{\src}{{\textit Swift} J1858.6$-$0814 }

\newcommand{\xmm}{\textit{XMM--Newton}\ }

\newcommand{\angstrom}{\mbox{\normalfont\AA}}

\def\aap{A\&A}

\def\aj{AJ}
\def\apj{ApJ}

\def\apjl{ApJ}

\def\mnras{MNRAS}
\def\nat{Nature}

\def\aapr{A\&ARev}

\usepackage{lineno}

\title{\center A persistent ultraviolet outflow from an accreting neutron star transient}

\author[1,$\star$]{N.~Castro~Segura}
\author[1]{C.~Knigge}
\author[2,3]{K.S.~Long}
\author[1]{D.~Altamirano}
\author[4,5]{M.~Armas~Padilla}
\author[6]{C.~Bailyn}
\author[7]{D.A.H.~Buckley}
\author[1]{D.J.K.~Buisson}
\author[4,5]{J.~Casares}
\author[1]{P.~Charles}
\author[8,23]{J.A.~Combi}
\author[4,5]{V.A.~C\'uneo}
\author[9]{N.D.~Degenaar}
\author[8]{S.~del~Palacio}
\author[10]{M.~D\'iaz Trigo}
\author[11]{R.~Fender}
\author[1]{P.~Gandhi}
\author[1]{M.~Georganti}
\author[1,12,13]{C.~Guti\'errez}
\author[14]{J. V.~Hernandez~Santisteban}
\author[15,16]{F.~Jim\'enez-Ibarra}
\author[17]{J.~Matthews}
\author[18]{M.~M\'endez}
\author[1]{M.~Middleton}
\author[4,5]{T.~Mu\~noz-Darias}
\author[1]{M.~\"{O}zbey Arabac{\i}}
\author[1,19]{M.~Pahari}
\author[11]{L.~Rhodes}
\author[9,20]{T.D.~Russell}
\author[21]{S.~Scaringi}
\author[9,11]{J.~van~den~Eijnden}
\author[6,22]{G.~Vasilopoulos}
\author[1]{F.M.~Vincentelli}
\author[1]{P.~Wiseman}

\affil[1]{Department of Physics \& Astronomy. University of Southampton, Southampton SO17 1BJ, UK}
\affil[2]{Space Telescope Science Institute, 3700 San Martin Drive, Baltimore, MD 21218, USA}
\affil[3]{Eureka Scientific, Inc. 2452 Delmer Street, Suite 100, Oakland, CA 94602-3017, USA}
\affil[4]{Instituto de Astrofísica de Canarias, E-38205 La Laguna, Tenerife, Spain}
\affil[5]{Departamento de Astrofísica, Universidad de La Laguna, E-38206 La Laguna, Tenerife, Spain}
\affil[6]{Department of Astronomy, Yale University, PO Box 208101, New Haven, CT 06520-8101, USA}
\affil[7]{South African Astronomical Observatory, PO Box 9, Observatory 7935, Cape Town, South Africa}
\affil[8]{Instituto Argentino de Radioastronom\'{\i}a (CONICET;CICPBA;UNLP), C.C. No 5, 1894, Villa Elisa, Argentina}
\affil[9]{Anton Pannekoek Institute for Astronomy, University of Amsterdam, Science Park 904, 1098 XH Amsterdam, The Netherlands}
\affil[10]{ESO, Karl-Schwarzschild-Strasse 2, D-85748 Garching bei M\"unchen, Germany}
\affil[11]{Astrophysics, Department of Physics, University of Oxford, Denys Wilkinson Building, Keble Road, Oxford OX1 3RH, UK}
\affil[12]{Finnish Centre for Astronomy with ESO (FINCA), FI-20014 University of Turku, Finland}
\affil[13]{Tuorla Observatory, Department of Physics and Astronomy, FI-20014 University of Turku, Finland}
\affil[14]{SUPA School of Physics \& Astronomy, University of St Andrews, North Haugh, St Andrews KY16 9SS, UK}
\affil[15]{Department of Physics and Astronomy, Macquarie University, Sydney, NSW 2109, Australia}
\affil[16]{Research Centre in Astronomy, Astrophysics and Astrophotonics, Macquarie University, Sydney, NSW 2109, Australia}
\affil[17]{Institute of Astronomy, University of Cambridge, Madingley Road, Cambridge CB3 0HA, UK}
\affil[18]{Kapteyn Astronomical Institute, University of Groningen, P.O. BOX 800, 9700 AV Groningen, The Netherlands}
\affil[19]{Department of Physics, IIT Hyderabad, Hyderabad 502285, India}
\affil[20]{INAF, Istituto di Astrofisica Spaziale e Fisica Cosmica, Via U. La Malfa 153, I-90146 Palermo, Italy}
\affil[21]{Centre for Extragalactic Astronomy, Department of Physics, Durham University, DH1 3LE, UK}
\affil[22]{Universit\'e de Strasbourg, CNRS, Observatoire astronomique de Strasbourg, UMR 7550, F-67000 Strasbourg, France}
\affil[23]{Departamento de Ingeniería Mecánica y Minera (EPSJ), Universidad de Jaén, Campus Las Lagunillas s/n, A3, E-23071 Jaén, Spain}
\affil[$\star$]{e-mail address: N.Castro-Segura@soton.ac.uk}

\begin{document}
\flushbottom
\maketitle
\clearpage
%
%

{\bf
\noindent All disc-accreting astrophysical objects produce powerful outflows. In binaries containing neutron stars (NS) or
black holes, accretion often takes place during violent outbursts.
The main disc wind signatures during these eruptions
are blue-shifted X-ray absorption lines, which are preferentially seen
in disc-dominated "soft states"\cite{Ponti2012MNRAS.422L..11P,Homan2016ApJ...830L...5H}. By contrast, optical wind-formed
lines have recently been detected in "hard states", when a hot corona
dominates the luminosity\cite{1858_optical_winds}. The relationship between
these signatures is unknown, and no erupting system 
has revealed wind-formed lines between
the X-ray and optical bands yet, despite the many strong resonance
transitions in this ultraviolet (UV) region\cite{LongKnigge2002ApJ...579..725L}. Here, we show that the
transient NS binary \src exhibits wind-formed,
blue-shifted absorption associated with C~{\sc iv}, N~{\sc v}
and He~{\sc ii} in time-resolved UV spectroscopy during a luminous
hard state. This represents the first evidence for a warm, 
moderately ionized outflow component in this state. Simultaneously
observed optical lines also display transient blue-shifted absorption.
Decomposing the UV data into constant and variable components, the blue-shifted
absorption is associated with the former. This implies that the outflow is not connect to the luminous flares in the data. The joint presence of UV and optical wind features reveals a
multi-phase and/or stratified outflow from the
outer disc\cite{Waters:2021ApJ...914...62W}. This type of persistent mass
loss across all accretion states has been predicted by
radiation-hydrodynamic simulations\cite{higginbottom+:2020} and
helps to explain the shorter-than-expected outbursts duration\cite{Tetarenko2018Natur.554...69T}.
}
\vspace{40pt}

On October 2018, the Neil Gehrels Swift Observatory ({\it Swift}\cite{SwiftTelescope2004ApJ...611.1005G}) detected a bright new X-ray binary transient, \src (hereafter J1858)\cite{J1858Krimm2018ATel12151....1K}. Multi-wavelength observations quickly led to the discovery of radio, optical and near-ultraviolet (UV) counterparts\cite{J1858AmiBright2018ATel12184....1B,J1858Vasilopoulos2018ATel12164....1V,J1858UVOTKennea2018ATel12160....1K}. The detection of thermonuclear runaway explosions in X-rays (Type~I X-ray bursts) established that the accreting object is a neutron star located at a distance of about $13$~kpc\cite{1858X-rayBursts2020MNRAS.499..793B}. The system was also found to undergo eclipses, implying a nearly edge-on viewing angle with respect to the disc ($i \gtrsim 70^{\circ}$) and revealing the orbital period to be $P_\mathrm{orb} \simeq 21.3$~h\cite{1858_Eclipses2021MNRAS.503.5600B}.

J1858 displayed extreme variability during its outburst in all energy bands, with the X-ray luminosity changing by 1--2 orders of magnitude on time-scales of seconds (see Figure~\ref{fig:LCs}-a)\cite{1858_Eclipses2021MNRAS.503.5600B,J1858Vasilopoulos2018ATel12164....1V}. 
The X-ray spectrum consisted of a heavily absorbed thermal accretion disc component plus a very shallow non-thermal power law tail (photon flux $N_{ph}(E) \propto E^{-\Gamma}$, with $\Gamma <1$) \cite{J1858NicerLudlam2018ATel12158....1L}. 
Both the peculiar X-ray spectrum and spectacular variability are reminiscent of those seen during the outbursts of the well-studied black-hole X-ray binaries {\em V404 Cyg} and {\em V4641  Sgr}, which are thought to be a consequence of accretion at super-Eddington rates\cite{ MottaV404Cyg_b2017MNRAS.471.1797M,J1858NusHare2020ApJ...890...57H}. 

In order to shed light on the accretion and outflow processes associated with the outburst, we carried out strictly simultaneous, time-resolved observations across the electromagnetic spectrum on August 6, 2019 around 00 (UTC). One of our primary goals was to search for outflow signatures in the far-ultraviolet (far-UV) band, since this region contains several strong resonance lines that are very sensitive to the presence of warm, moderately-ionized intervening material.Therefore, the timing of this campaign was centered on far-UV spectroscopic observations with the Hubble Space Telescope (HST). Simultaneous optical spectroscopy was obtained at both the Very Large Telescope (VLT) array and the Gran Telescopio de Canarias (GTC). Additional information about the campaign is provided in the Extended Data section.

In line with data obtained at other wavelengths\cite{J1858Vasilopoulos2018ATel12164....1V,1858_radio_Eijnden2020MNRAS.496.4127V}, the far-UV light curve exhibits dramatic flaring activity (Figure~\ref{fig:LCs}-b). The X-ray, far-UV and optical variability are clearly correlated, with any lags between these time series being $\lesssim 3$~s (Vincentelli et al. in prep). In agreement with previously  reported X-ray light curves\cite{1858X-rayBursts2020MNRAS.499..793B,1858_Eclipses2021MNRAS.503.5600B}, there is no evidence of thermonuclear runaway events during the campaign. This suggests that the multi-wavelength flaring is driven by a variable central X-ray source.

\begin{figure}[ht]
\centering
\includegraphics[width=1\linewidth]{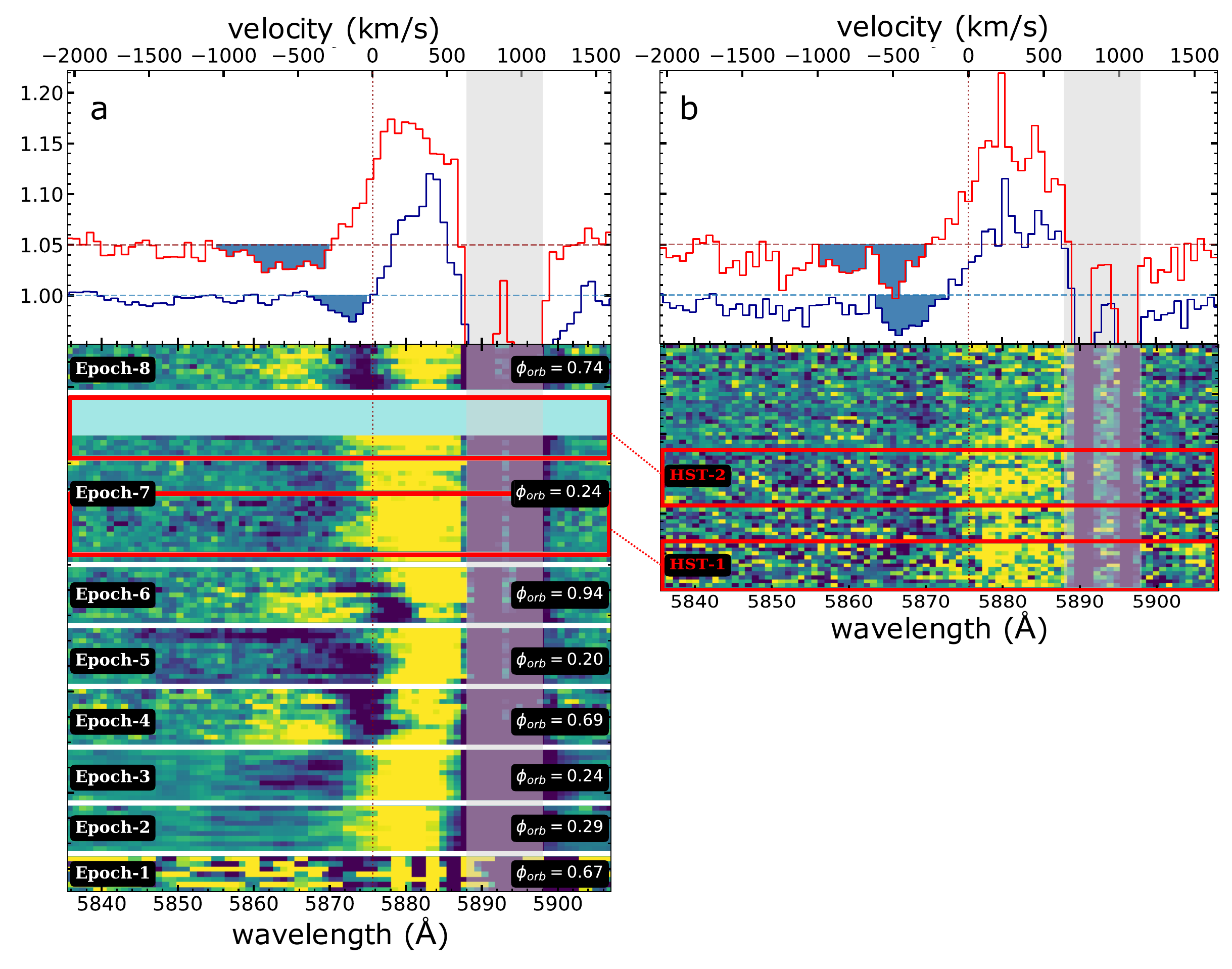}
\caption{{\bf Apparently transient optical wind signatures.} (a) Trailed spectra of the 8 {\em GTC/OSIRIS} epochs published in Mu\~noz-Darias et al. (2020)\cite{1858_optical_winds} with the corresponding orbital phase\cite{1858_Eclipses2021MNRAS.503.5600B} and (b) {\em VLT/X-Shooter} during the HST visit centered on He~{\sc i}~5876~\AA\,. The average spectrum of all the observations is shown in the top panel with a blue line. Strictly simultaneous observations during the two ultraviolet exposures are highlighted in red boxes, with their corresponding averaged spectrum shown in red in the top panels with a 5\% offset for clarity. Absorption troughs below continuum levels are highlighted with a shaded area. Telluric absorption region around $\lambda~5836$~\AA\ is indicated with the shaded vertical band.
}
\label{fig:trailed}
\end{figure}

The presence of a large, strongly irradiated accretion disc is the key requirement for a thermally-driven outflow \cite{Begelman1983ApJ...271...70B,Done2007AARv..15....1D,Higginbottom2015ApJ...807..107H}, while high inclinations tend to strengthen wind-formed absorption features \cite{Ponti2012MNRAS.422L..11P,higginbottom+:2020}. All of this makes J1858 an ideal candidate for displaying clear observational outflow signatures. As summarised in the {\em Methods} section, X-ray spectroscopy of the source obtained earlier in the same outburst already found tentative evidence for an outflow\cite{1858X-rwinds2020MNRAS.498...68B}. Time-resolved optical spectroscopy also revealed clear, but highly variable P~Cygni wind features in H$\alpha$ and He~{\sc i}~5876~\AA\ during the bright hard state \cite{1858_optical_winds} (Figure~\ref{fig:trailed}).

\begin{figure}[ht]
\centering
\includegraphics[width=\linewidth]{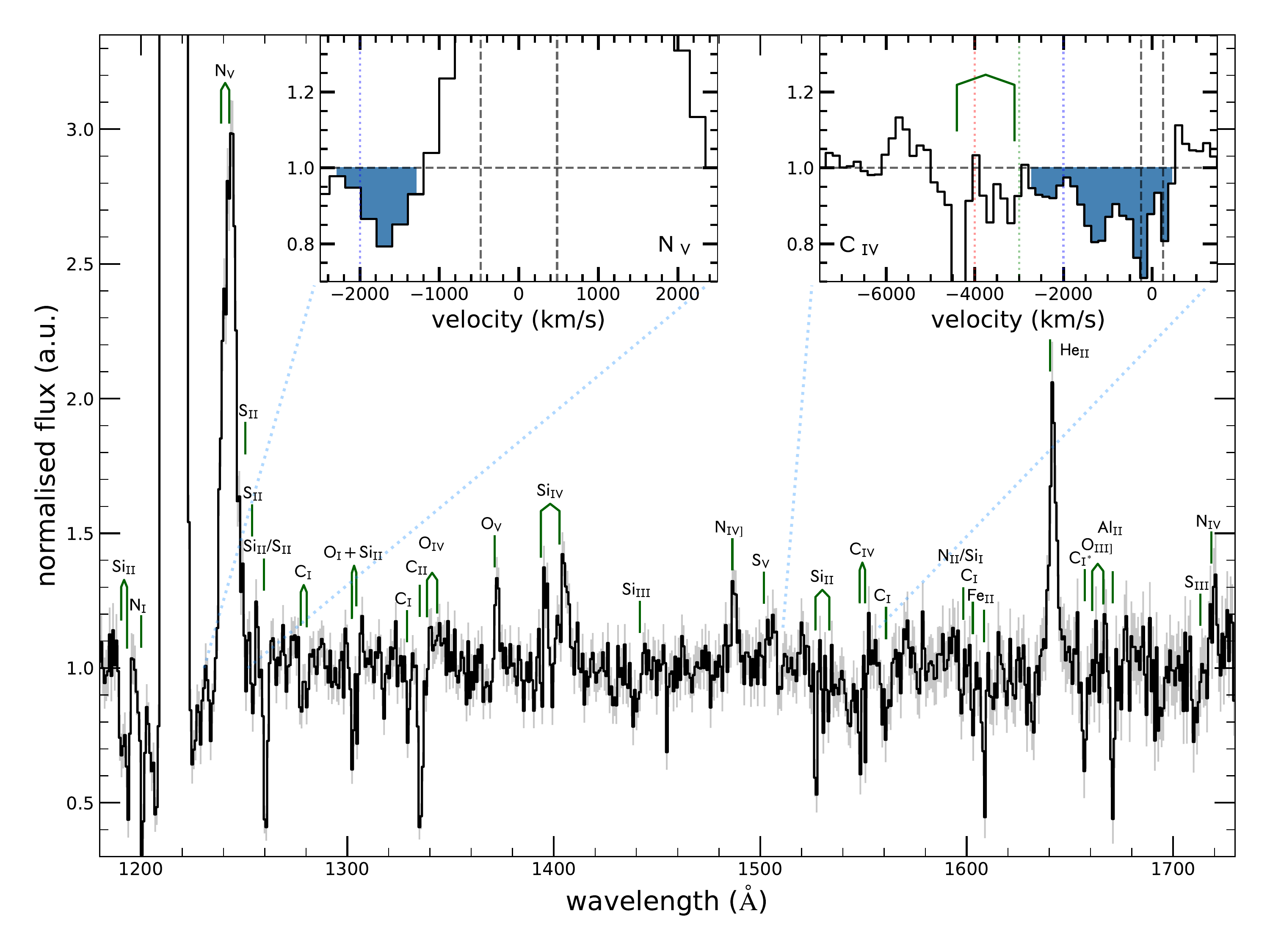}
\caption{{\bf Average far-UV spectrum of \src during the luminous hard state.} Numerous emission and absorption lines are present; the dominant transitions have been labeled with their corresponding rest position indicated with a green tick. All the emission components are skewed toward shorter wavelengths with blue absorption troughs, which are the characteristic footprint of disc outflows. Light gray area represent the standard error. Insets show a zoom-in to the N~{\sc v} ($\lambda\lambda1284-1437$~{\AA}) and C~{\sc iv} ($\lambda\lambda1513-1668$~{\AA}) profiles with the blue-shifted absorption signatures highlighted in blue, in the latter nearby Si~{\sc ii} interstellar absorption is indicated with connected green ticks. These signatures indicate the presence of a warm, moderately ionized accretion disc wind with characteristic velocities similar to those observed in the optical.}
\label{fig:wind}
\end{figure}

Figure~\ref{fig:wind} shows the time-averaged far-UV spectrum we obtained with HST in the hard state. The spectrum is rich in both absorption and emission lines that span a wide range of ionization states. Most of the low-ionization absorption lines are centered at or near the rest wavelength of the relevant transition, with most of these lines not being intrinsic to the system but rather due to interstellar absorption along the line of sight. However, at least two emission lines -- N~{\sc v}~1240~\AA\ and C~{\sc iv}~1550~\AA\ -- show clear evidence for associated blue-shifted absorption. Since these species are associated with temperatures of $T \simeq \mathrm{a~few} \times 10^{4}$~K, their presence unambiguously establishes the existence of a warm and moderately ionized outflowing component. 

Several other transitions -- e.g. O~{\sc v}~1370~\AA\ and Si~{\sc iii}~1440~\AA\ -- also show tentative evidence for blue-shifted absorption. Moreover, all strong emission lines in the spectrum -- which includes the Si~{\sc iv}~1400\AA\ doublet resonance line and the He~{\sc ii}~1640~\AA\ recombination feature -- show evidence for a slight red-shift or a red-skew, suggesting that they are also affected by blue-shifted wind absorption. 

As shown in the insets of Figure~\ref{fig:wind}, the blue edges of the far-UV absorption features extend up to $\simeq -2000~\mathrm{km~s^{-1}}$, similar to the wind speed inferred from the optical data. However, the far-UV absorption troughs are considerably deeper than those in the optical, which rarely fall below $90\% - 95\%$ of the continuum. This is likely because most of the strong far-UV lines are associated with strongly scattering resonance transitions, whereas the optical features are associated with recombination lines that connect two excited levels. Very high (column) densities are required in order for such recombination lines to produce absorption. On the other hand, sensitivity of far-UV resonance lines to intervening material makes this waveband particularly valuable for studying outflows.

It is important to establish if the far-UV wind signatures are always present or are instead associated with the strong flaring events in the light curve. We have therefore carried out a maximum likelihood linear decomposition of the time-resolved spectroscopy into a constant and a flaring (variable) component. The spectra inferred for the two components are shown in Figure~\ref{fig:varfit}; details regarding the decomposition technique are provided in the {\em Methods} section. In both N~{\sc v}~1240~\AA\ and C~{\sc iv}~1550~\AA, the blue-shifted absorption signature is clearly associated with the constant component. This suggests that either our line of sight to the emitting region responsible for the flaring component does not pass through the warm outflow or that the ionization state of the outflow changes significantly during the flares. 
Perhaps more importantly, it also suggests that the outflow is, in fact, always present, but that its signatures may sometimes be swamped by the flaring component (in which these signatures are absent). The same effect may be responsible for the transience of the  blue-shifted absorption seen in the optical data, especially considering how weak these features are (c.f. Figure~\ref{fig:trailed}).

The presence of detectable blue-shifted absorption associated with the UV resonance lines (e.g. N~{\sc v}~1240~\AA, C~{\sc iv}~1550~\AA) implies that the optical depth in these transitions must be significant. This, in turn, requires minimum column densities for the relevant ions, which can be cast as approximate lower limits on the mass-loss rate carried away by the outflow (see {\em Methods} for details). Conservatively assuming ionization fractions of $f=1$ for both $C^{3+}$ and $N^{4+}$, these limits are $\dot{M}_{wind} \gtrsim 2 \times 10^{-11}~{\mathrm M_{\odot}~yr^{-1}}$ for N~{\sc v}~1240~\AA\ and $\dot{M}_{wind} \gtrsim 3 \times 10^{-12}~{\mathrm M_{\odot}~yr^{-1}}$ for C~{\sc iv}~1550~\AA . The actual ionization fractions may be considerably lower, and the mass-loss rate correspondingly higher. 

The apparent time-averaged X-ray luminosity during the flaring hard state in which we observed J1858 was $L_{X} \simeq 0.01 L_\mathrm{Edd}$, although individual flares appear to have reached super-Eddington levels \cite{1858X-rayBursts2020MNRAS.499..793B}. Taken at face value, this corresponds to an average accretion rate in this state of $\dot{M}_\mathrm{acc} \simeq 10^{-10}~\mathrm{M_{\odot}~yr^{-1}}$. In this case, $\dot{M}_\mathrm{wind} / \dot{M}_\mathrm{acc} \gtrsim 0.2$, suggesting that the wind is dynamically important and could significantly affect the accretion flow \cite{V404Cyg_opt_winds2016Natur.534...75M,Shields1986ApJ...306...90S}\textsuperscript{, but also see }\cite{ganguly}. However, it is also possible that the intrinsic luminosity was much higher throughout this state, with time-variable obscuration being responsible for the reduction in the time-averaged flux (and perhaps also the flaring activity). Such obscuration need not necessarily be associated with the disk wind itself (see {\em Methods})

In the extreme case that $L \simeq L_{Edd}$, the constraint on the wind efficiency is $\dot{M}_\mathrm{wind} / \dot{M}_\mathrm{acc} \gtrsim 10^{-3}$. 

The discovery of optical, UV and (probably) X-ray outflow signatures in the luminous hard state of J1858 suggests that disc winds may {\em always} be present in transient X-ray binaries, not just in disc-dominated soft states. Our identification of the 
{\em constant} (non-flaring) spectral component as the carrier of these signatures in the far-UV strongly supports this idea. 
X-ray and far-UV wind signatures have also been observed in some {\em persistent} soft-state X-ray binaries,\cite{Ioannou+2003AA...399..211I,Neilsen2020ApJ...902..152N}, i.e. systems in which the disc is not subject to the instability that drives the outbursts of transient accretors \cite{Dubus2019AA...632A..40D}.

The emerging physical picture of disc winds being an integral part of the accretion flows in X-ray binaries is consistent with theoretical modeling of outburst light curves\cite{Dubus2019AA...632A..40D,Tetarenko2018Natur.554...69T}. It is also in line with radiation-hydrodynamical modeling of thermally-driven outflows from X-ray binary discs\cite{luketic,done18,higginbottom+:2020}. These simulations confirm that strong mass loss is inevitable in any systems with a sufficiently large disc subject to strong irradiation, regardless of accretion state\cite{higginbottom+:2020}. These conditions are met in J1858 (see {\em Methods}). A key test of the thermally-driven wind scenario will be to check that wind signatures are absent in systems where these conditions are not met\cite{charles_shortperiod_wind}.

Regardless of the driving mechanism, two key outstanding questions are where and how these outflows manage to sustain a sufficiently low ionization state to allow the formation of optical and UV lines. The most likely answers are that self-shielding, probably coupled with clumping, protects parts of the dense base of the wind above the outermost disc regions from over-ionization. Indeed, 
recent hydrodynamical simulations of irradiated discs in active galactic nuclei predict the existence of clumpy, thermally unstable, multi-phase outflows\cite{Waters:2021ApJ...914...62W}. This mechanism might also be at work in X-ray binaries, but new  radiation-hydrodynamic simulations will be needed to confirm this. 


\subsection*{Acknowledgments}

NCS \& CK acknowledge support by the Science and Technology Facilities Council (STFC), and from STFC grant ST/M001326/1. Partial support for KSL's effort on the project  was provided by NASA through grant numbers HST-GO-15984 and HST-GO-16066 from the Space Telescope Science Institute, which is operated by AURA, Inc., under NASA contract NAS 5-26555. NCS thanks Tricia Royle for helping coordinate the time-critical observations in this article. ND acknowledges support from a Vidi grant for the Netherlands Organization for Scientific Research (NWO). JVHS acknowledges support from STFC grant ST/R000824/1. MAP, JC, FJI and TMD acknowledge support from the grant AYA2017-83216-P and PID2020-120323GB-I00. TMD also acknowledges RYC-2015-18148 and EUR2021-122010. TMD and MAP acknowledge support from grants with references ProID2020010104 and ProID2021010132. JM acknowledges a Herchel Smith Fellowship at Cambridge. TDR acknowledge financial contribution from the agreement ASI-INAF n.2017-14-H.0. JvdE is supported by a Lee Hysan Junior Research Fellowship from St Hilda’s College, Oxford. GV acknowledges support by NASA Grants 80NSSC20K1107, 80NSSC20K0803 and 80NSSC21K0213.  M\"{O}A acknowledges support from the Newton International Fellowship program from Royal Society. JAC acknowledge from grants PICT-2017-2865 (ANPCyT), PID2019-105510GB-C32/AEI/10.13039/501100011033, FQM-322, as well as FEDER funds.

\subsection*{Author contributions statement}
NCS and CK wrote the original proposal, performed the data analysis and wrote the paper with significant feedback from 
KSL, DA, FMV, SdP, JM and MM. 
MAP JC FJI and TMD provided the GTC data. 
JVHS reduced the X-Shooter data and assisted in designing the observations.
DJKB and DA provided the X-ray data. 
DAHB, JAC, VAC, NDD, SdP, MDT, RF, CG, JVHS, MP, M\"{O}A, LR, TDR, JvdE, FMV and PW assisted in proposing and planning the MW observations. 
All authors contributed to the original proposal, discussed the results and commented on the manuscript.

\subsection*{Author information}
Reprints and permissions information is available at www.nature.com/reprints. The authors declare no competing financial interests. All correspondence should be addressed to N.C.S. (N.Castro-Segura@Soton.ac.uk)

\begin{figure}[ht]
\centering
\includegraphics[width=0.85\linewidth]{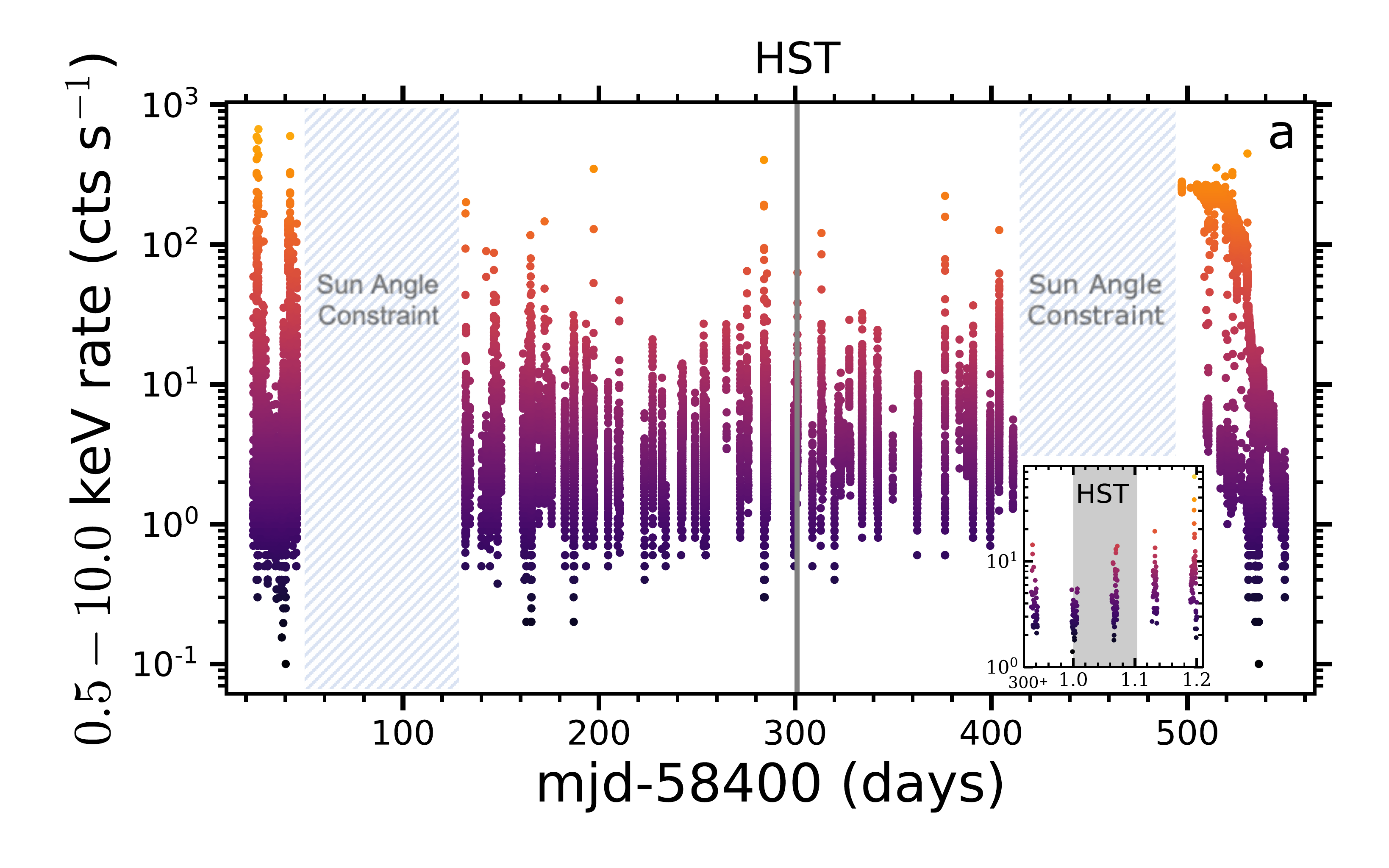}\\
\center\includegraphics[width=0.85\linewidth]{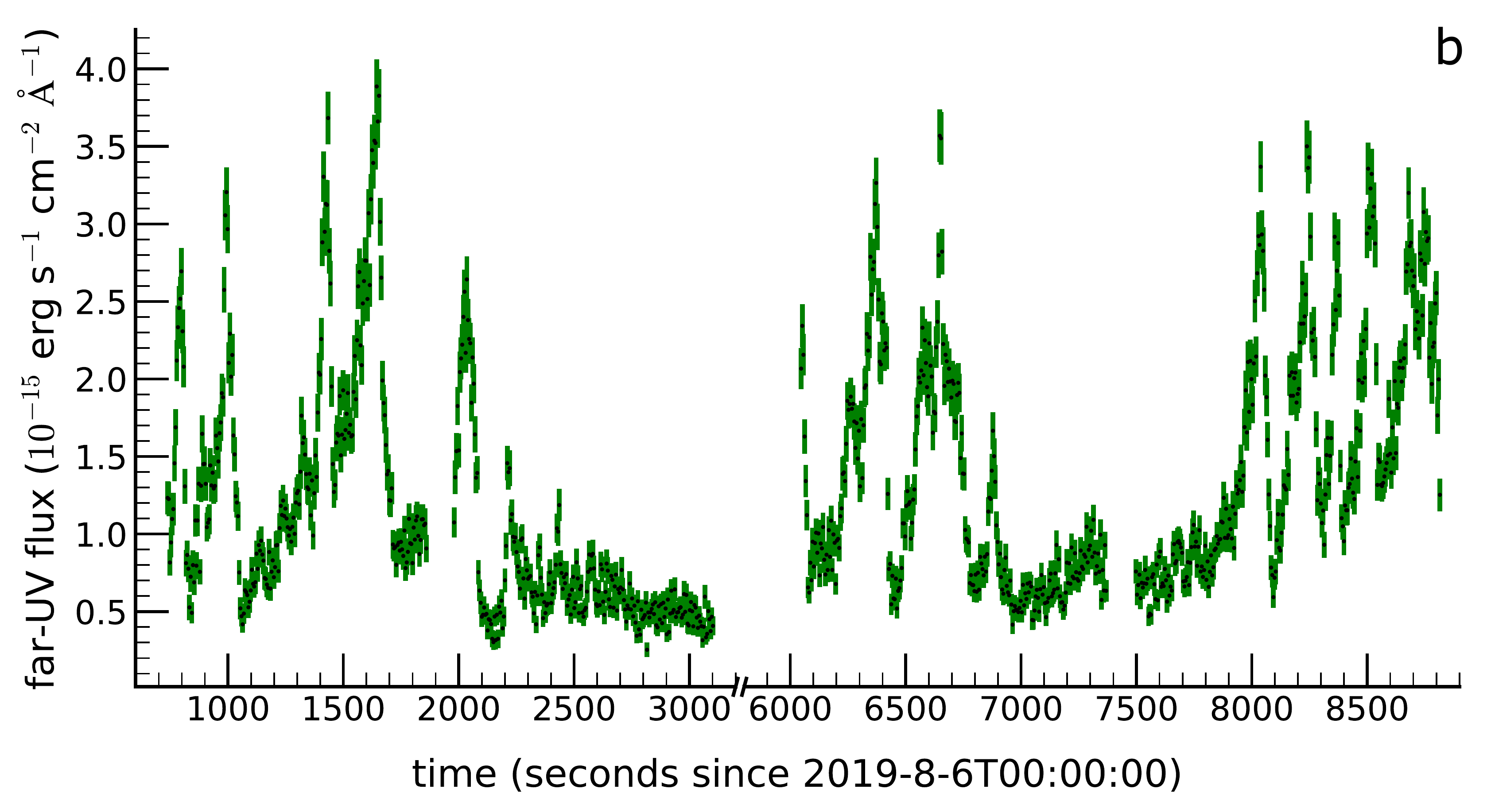}
\caption{{\bf Overview light curves of the X-ray transient \src}.
(a) Outburst evolution as observed 
with {\em NICER} in the 0.5--10keV band (reproduced from Buisson et al. 2020\cite{1858_Eclipses2021MNRAS.503.5600B}); the two large gaps are caused by Sun constraints. 
The source exhibits flares that reach the Eddington limit during the first 450 days while it is in the hard state. 
The time of the HST far-UV observations is marked with a vertical line. Colour code refers to the observed count rate. Inset shows a zoom-in around the time of the HST observations (MJD$\simeq 58701$), indicated with the shaded area. The inset cover the region around the time of the HST visit.
(b) HST far-UV light-curve in 5s bins, showing strong flares (up to a factor of 10 increase in flux) and flickering at lower flux levels. Green bars represent the standard error, the mean is indicated with small black dot.}
\label{fig:LCs}
\end{figure}

\begin{figure}[ht]
\centering
\includegraphics[width=\linewidth]{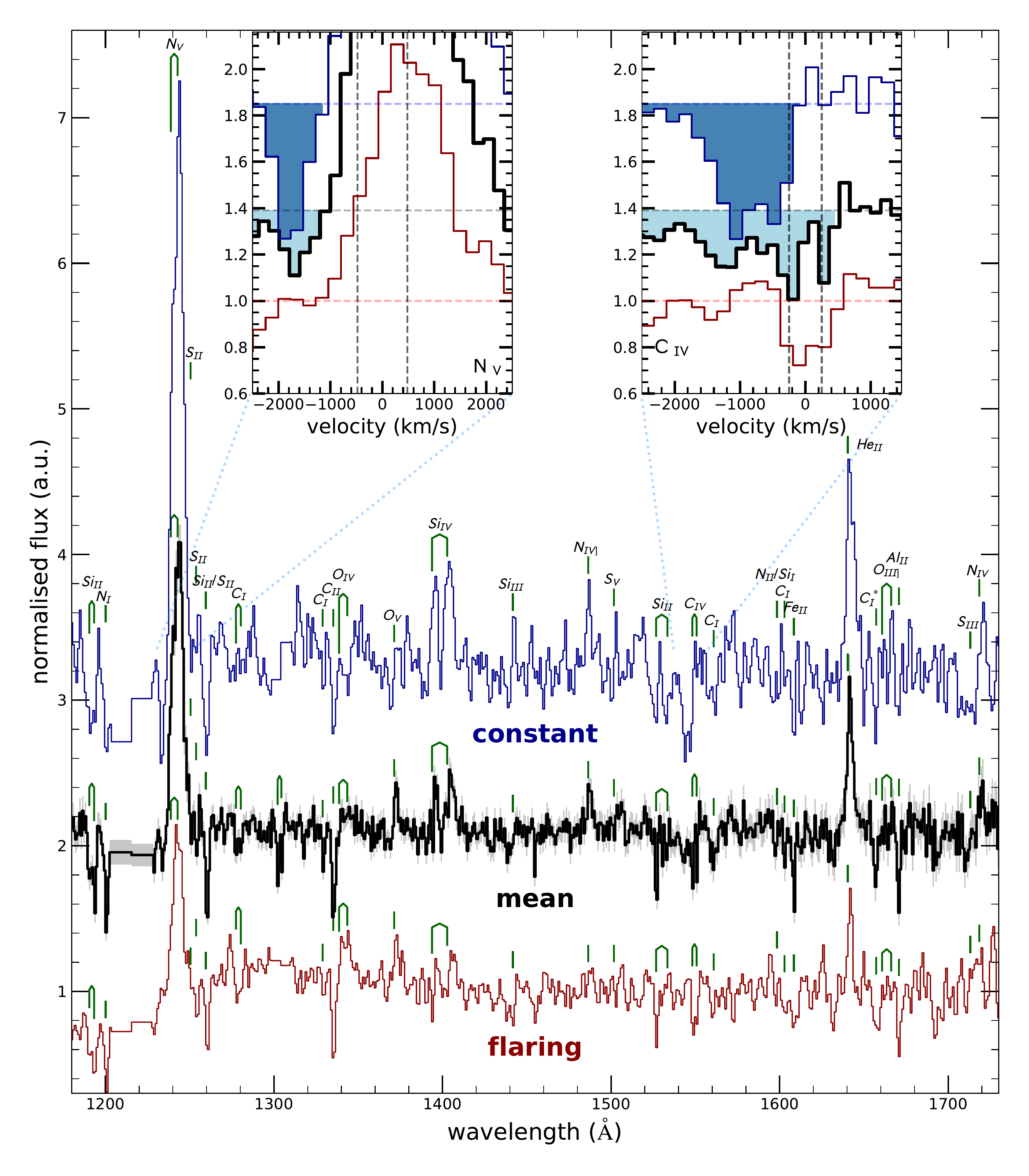}
\caption{{\bf Spectral decomposition into a constant (blue) and flaring component (red)}. Being the latter driven by the observed continuum variability in the far-UV. The average normalized spectrum is displayed with a thick black line for reference, and all are normalized to the continuum level. An offset has been added to the spectra for clarity. The regions of geocoronal emission like Lyman $\alpha$ and Si~{\sc ii} were removed to avoid artifacts in the spectral decomposition. Rest positions of the dominant ions are marked with a green tick and labeled in the top spectrum. Insets are zooms into the two transitions in which the presence of the outflow is more prominent. Specifically, regions covered in the insets are $\lambda\lambda1284-1474$~{\AA} for N~{\sc v} and $\lambda\lambda1525-1717$~{\AA} for C~{\sc iv}.}
\label{fig:varfit}
\end{figure}

\newpage
\section*{Methods}

\subsection*{Previous X-ray Spectroscopy}
\src was observed by \xmm with the {\em Reflection Grating Spectrometer} (RGS) during March/April of 2019, a few months before the multi-wavelength campaign discussed in the present paper. A detailed analysis of {\it XMM--Newton/RGS} observations has already been presented elsewhere\cite{1858X-rwinds2020MNRAS.498...68B}, so here we just briefly summarize some relevant results.

As during our campaign, \src was in a hard spectral state at the time of the {\it XMM--Newton/RGS} observations. In fact, there is no significant difference in the X-ray hardness ratios between the two epochs as seen by the all sky monitoring telescopes {\it MAXI} and {\it Swift/BAT}.

A search for wind signatures in the X-ray spectrum obtained with the {\it RGS} did not reveal any clear blue-shifted absorption features, which would be the "smoking gun" for the presence of outflowing material. However, the N~{\sc vii} emission line is significantly redshifted, which suggests that photons in the blue wing might be scattered out of our line of sight by an outflow. Thus, there is tentative evidence for an X-ray processing wind in the hard state from these observations. 
Assuming a neutron star with typical mass of $1.4 M_\odot$, the velocity dispersion observed in this narrow line indicates that the bulk of the emission is originated at $2-8 \times 10^{9}\ {\rm cm}$ from the central source, consistent with the wind being thermally driven discussed bellow. 

Finally, the absorbing column density inferred during its soft state is $N_H \simeq 2\times 10^{21}\ {\rm cm^2}$, similar to the galactic extinction in the line of sight\cite{1858X-rayBursts2020MNRAS.499..793B}.

\subsection*{Ultraviolet Spectroscopy}

\subsubsection*{Observations}

 \src was observed  under the program GO/DD 15984 with N. Castro Segura as PI on Aug 5, 2019 23:49:20 UT. A total of $4.9\ \mathrm{kilo~seconds\, (ks)}$ exposure was obtained in the far-UV with the {\em Cosmic Origin Spectrograph} (COS\cite{COS}) and the G140L grating using the primary science aperture (PSA). This configuration provides a spectral resolution of $\mathrm{R = \lambda / \Delta \lambda} \sim 900$. All the observations were obtained in TIME-TAG mode, yielding a stream of detected events at a time resolution of $32\, \mathrm{ms}$.
 
\subsubsection*{Data analysis}
We reduced the far-UV data using the HST {\sc calcos} pipeline\footnote{Provided by The Space Telescope Science Institute (\url{https://github.com/spacetelescope})}. One-dimensional spectra were extracted using the {\sc twozone} algorithm, which sums over the cross-dispersion direction such that 99\% of the flux is extracted at each wavelength. Errors are estimated from Poisson statistics, and the background is modeled with a smooth polynomial and subtracted from the target spectrum. 
We extracted light curves from the {\sc time-tag} event files using the same regions defined by the pipeline, except that empirical background correction was directly applied. Regions affected by geocoronal airglow emission associated with $Lyman~\alpha$ ($\lambda\lambda 1208-1225$~{\AA}) and O~{\sc ii} ($\lambda\lambda 1298-1312$~{\AA}) were masked when extracting the light curves.

\subsubsection*{Spectral decomposition}

The highly variable far-UV luminosity during our observations gives rise to a strongly bimodal logarithmic flux distribution (Extended Data Figure~\ref{fig:lc_hist}). This is in line with the visual impression from the far-UV light curve that the dominant variability is due to "shots" or "flares" superposed on a roughly constant background (Figure~\ref{fig:LCs}-b). 

We have isolated the spectra associated with these two components by carrying out a simple linear decomposition of our time-resolved spectroscopic data into a constant and a flaring (variable) component. Following Eracleous \& Horne (1996)\cite{EracleousHorne:1996}, we assume that the flux density $F(\lambda, t)$ at wavelength $\lambda$ and time $t$ can be written as 

\begin{equation}
\label{eq:decomp}    
F(\lambda,t) = C(\lambda) + V(\lambda) D(t),
\end{equation}
where $C(\lambda)$ and $V(\lambda)$ are the spectra of the constant and flaring components, respectively. The function $D(t)$ is the driving light curve of the flaring component. 

In order to estimate $D(t)$, we constructed a far-UV continuum light curve at 10~s time resolution. We then estimated the underlying constant level in this light curve and created a normalized driving light curve from which this estimate was removed. 
We finally smooth the resulting time series with a 5-point, second-order Savitzky-Golay filter to obtain our estimate of $D(t)$. The result is shown as the red curve in Extended Data Figure~\ref{fig:fdrive}. 

With $D(t)$ fixed, the decomposition described by Equation~\ref{eq:decomp} becomes just a series of $N_\lambda$ 2-parameter fits, where $N_\lambda$ is the number of wavelength bins being considered. Since {\em HST/COS} uses a photon-counting far-UV detector, the data set actually consists of a time- and wavelength-tagged event stream. Our decomposition is therefore based an unbinned (in time) maximum likelihood fit to the data at the individual photon-event level, since this maximizes the signal-to-noise ratio of the inferred spectral components. The log-likelihood for this model can be derived from Poisson statistics and turns out to be
\begin{equation}
\label{eq:loglike}    
\ln{\mathcal{L}} \propto \left[\sum_{i=0}^{N_{phot}} \ln\left[\mathcal{C}(\lambda) + D(t)\mathcal{V}(\lambda)\right]\right] - \mathcal{N}.
\end{equation}
Here, $\mathcal{C}$ and $\mathcal{V}$ are now the count-rates associated with the constant and variable components, and $\mathcal{N}$ is the total number of photons predicted by the model. We obtain best-fit estimates of $\mathcal{C}$ and $\mathcal{V}$ by maximising Equation~\ref{eq:loglike}. The flux-calibrated spectra described as constant and flaring components in Figure~\ref{fig:varfit} are then constructed in the usual way, by multiplying by the wavelength-dependent inverse sensitivity curve. Detector regions dominated by background and/or geo-coronal emission are excluded from the fit. 

\subsection*{Optical spectroscopy}

During the {\em HST} visit, strictly simultaneous observations of \src were carried out with {\em X-Shooter}\cite{XShooter} (program ID 2103.D-5052(A)) and {\em OSIRIS}\cite{OSIRIS} spectrographs (program ID GTC23-19A), mounted on the {\em VLT UT2-Kueyen} telescope in Paranal Observatory and in {\em GTC} at Roque de los Muchachos Observatory, respectively. {\em X-Shoooter} yielded time-resolved optical/NIR spectra covering the range $\lambda\lambda0.3-2.4~\mu {\rm m}$. With this instrument we obtained a total of 58 individual exposures with integration times of $\simeq 300~\mathrm{s}$, for a total exposure time of $\simeq 17.4~\mathrm{ks}$. We used slit widths of 0.9 and 1.0 arcsec in the UVB and visible (VIS), respectively, yielding corresponding velocity resolutions of $\sim 51,
~33~\mathrm{km~s^{-1}}$. The data set was reduced using the standard ESO pipeline {\sc EsoReflex}\cite{ESOREFLEX} version 3.3.5. Calibration frames were taken every 1~h and additionally during the occultation of HST by the Earth. A total of 20 science exposures of five minutes long were gathered with {\em GTC/OSIRIS}, covering the first two hours of the campaign, using the grism R2500R ($\lambda\lambda$5575 -- 7685~{\AA}) and one with R1000B ($\lambda\lambda$4200 -- 7400~{\AA}), delivering a velocity resolution of $\sim 160~\mathrm{km~s^{-1}}$ and $\sim350~\mathrm{km~s^{-1}}$ respectively. Further details on the data reduction of these observation are given in Mu\~noz-Darias et al. (2020)\cite{1858_optical_winds}.

\subsubsection*{Outflow diagnostics}

The presence of blue-shifted absorption associated with far-UV and optical lines implies a significant column density of material in the lower level of the relevant atomic transition. This, in turn, can be used to set a rough lower limit on the mass-loss rate of the outflow.

Following \cite{Drew:1987}, we approximate the outflow as spherical and adopt a simple Hubble-like $v \propto R$ velocity law. Combining the expression for the Sobolev optical depth with the continuity equation, the characteristic optical depth presented by such an outflow at velocity $v$ in a given line can be written as

\begin{equation}
\tau \simeq 74.1 
\left(\frac{f_\mathrm{osc}}{0.2847}\right)
\left(\frac{\lambda}{1549.062~\angstrom}\right)
\left(\frac{A}{7\times 10^{-5}}\right)
\left(\frac{f_\mathrm{ion}}{1.00}\right)
\left(\frac{\dot{M}_{w}}{10^{-10}~\mathrm{M_{\odot}~yr^{-1}}}\right)
\left(\frac{v}{1500~\mathrm{km~s^{-1}}}\right)^{-2}
\left(\frac{R(v)}{10^{10}~\mathrm{cm}}\right)^{-1}.
\label{eq:tau}
\end{equation}
Here, $f_\mathrm{osc}$ and $\lambda$ are the oscillator strength and wavelength of the line, respectively, $A$ is the abundance of the relevant element, $f_\mathrm{ion}$ is the fraction of those atoms in the correct ionization level, $\dot{M}_{w}$ is the mass-loss rate of the outflow, and $R(v)$ is the radius where velocity $v$ is reached in the wind. 

The reference values adopted for $f_\mathrm{osc}$, $\lambda$ and $A$ in Equation~\ref{eq:tau} are representative of the C~{\sc iv} resonance line (treated as a singlet). The reference velocity, $v \simeq 1500~\mathrm{km~s^{-1}}$, is chosen based on the location of the  blue-shifted absorption trough in the far-UV line profiles (cf Figure~\ref{fig:varfit}). Our adopted value of $R(v) \simeq 10^{10}~\mathrm{cm}$ corresponds to the radius in the disc beyond which a thermally driven outflow is expected to be launched (see below); it is also roughly the radius where $v_\mathrm{esc} \simeq 1500~\mathrm{km~s^{-1}}$. Finally, by taking $f_\mathrm{ion} = 1$, we ensure that our estimate of $\dot{M}_{w}$ is a lower limit (modulo uncertainties in the other parameters). 

Based on the depth of the absorption features in the far-UV line profiles, we expect that $\tau \gtrsim 1$ for both N~{\sc v} and C~{\sc iv}. The estimated lower limits on the mass-loss rates are then 
$\dot{M}_{w} \gtrsim 2 \times 10^{-11}$ from N~{\sc v} and $\dot{M}_{w} \gtrsim 3 \times 10^{-12}$ from C~{\sc iv}. 
The larger of these numbers corresponds to a Hydrogen column density of $N_H \simeq 2\times 10^{19}$cm$^{-2}$, if we adopt the same quasi-spherical wind model with an inner radius of $10^{10}$~cm. For comparison, a total column of 
$N_H \simeq 10^{24}$cm$^{-2}$ is required for the electron-scattering optical depth to reach $\tau_{es} \simeq 1$, as might be expected if the observed flaring is driven by time-dependent obscuration.

\subsection*{A thermally driven disc wind in \src ?}

The accretion discs in luminous X-ray binaries are subject to strong irradiation. As a result, the upper layers of the atmosphere can be heated to the inverse Compton temperature, which depends only on the spectral energy distribution of the radiation field. The X-ray spectrum of \src in the hard state can be approximated as a power law with photon index $\Gamma = 1.5$ and an exponential cut off at $E_\mathrm{max} \simeq 30$~keV\cite{1858X-rwinds2020MNRAS.498...68B}. 
For such a spectrum, the Compton temperature is approximately $kT_{IC} \simeq E_\mathrm{max} / 12$ \cite{Done2010arXiv1008.2287D}, which gives $T_{IC} \simeq 3 \times 10^{7}$~K for \src. 

Mass loss from these heated layers is inevitable at radii where the characteristic thermal speed of the ions, $v_\mathrm{th} \simeq 3 k T_{IC}  / m_p$ exceeds the local escape velocity, $v_\mathrm{esc} \simeq 2 G M / R$. Discs larger than the so-called Compton radius, $R_{IC} = (2 G M m_p)/(3 k T_{IC})$, are therefore expected to produce thermally driven outflows. For \src, we obtain $R_{IC} \simeq 5 \times 10^{10}$~cm. In reality, the radius at which this mechanism turns on is typically $R_\mathrm{min} \simeq 0.1~R_{IC}$ \cite{woods,pk02,done18}. In our mass-loss rate calculation above, we have adopted a characteristic radius $R \simeq 0.3 R_{IC}$ for the line-forming region in the outflow. 

The disc in \src is certainly large enough to drive such an outflow. The orbital period of the system is $P_\mathrm{orb} \simeq 21.3$~h \cite{1858_Eclipses2021MNRAS.503.5600B}. From Kepler's third law, and assuming that $q = M_2 / M_1 \lesssim 1$, the binary separation is $a_\mathrm{bin} \simeq 3 \times 10^{11}$~cm. If the disc is tidally limited, its outer radius will be roughly $R_\mathrm{disc} \simeq 0.9 R_1$, where $R_1$ is the Roche-lobe radius of the neutron star \cite{FrankKingRaine2002apa..book.....F}. The outer disc radius is therefore expected to be $R_\mathrm{disc} \simeq 1-2 \times 10^{11}{\rm cm}$ -- much larger than $R_{IC}$, let alone $R_\mathrm{min} \simeq 0.1~R_{IC}$.

The final condition for {\em strong} thermally driven mass loss is that the irradiating luminosity should be sufficiently strong, $L \gtrsim L_{crit} = 0.05 L_{Edd}$\cite{Begelman1983ApJ...271...70B}. This is comparable to the time-averaged luminosity in the flaring hard state of \src \cite{J1858NusHare2020ApJ...890...57H}. It is therefore likely that the system was luminous enough to drive a powerful thermal disc wind.

\renewcommand{\figurename}{Extended Data Figure}
\setcounter{figure}{0} 

\begin{figure}[ht]
\centering
\includegraphics[width=0.9\linewidth]{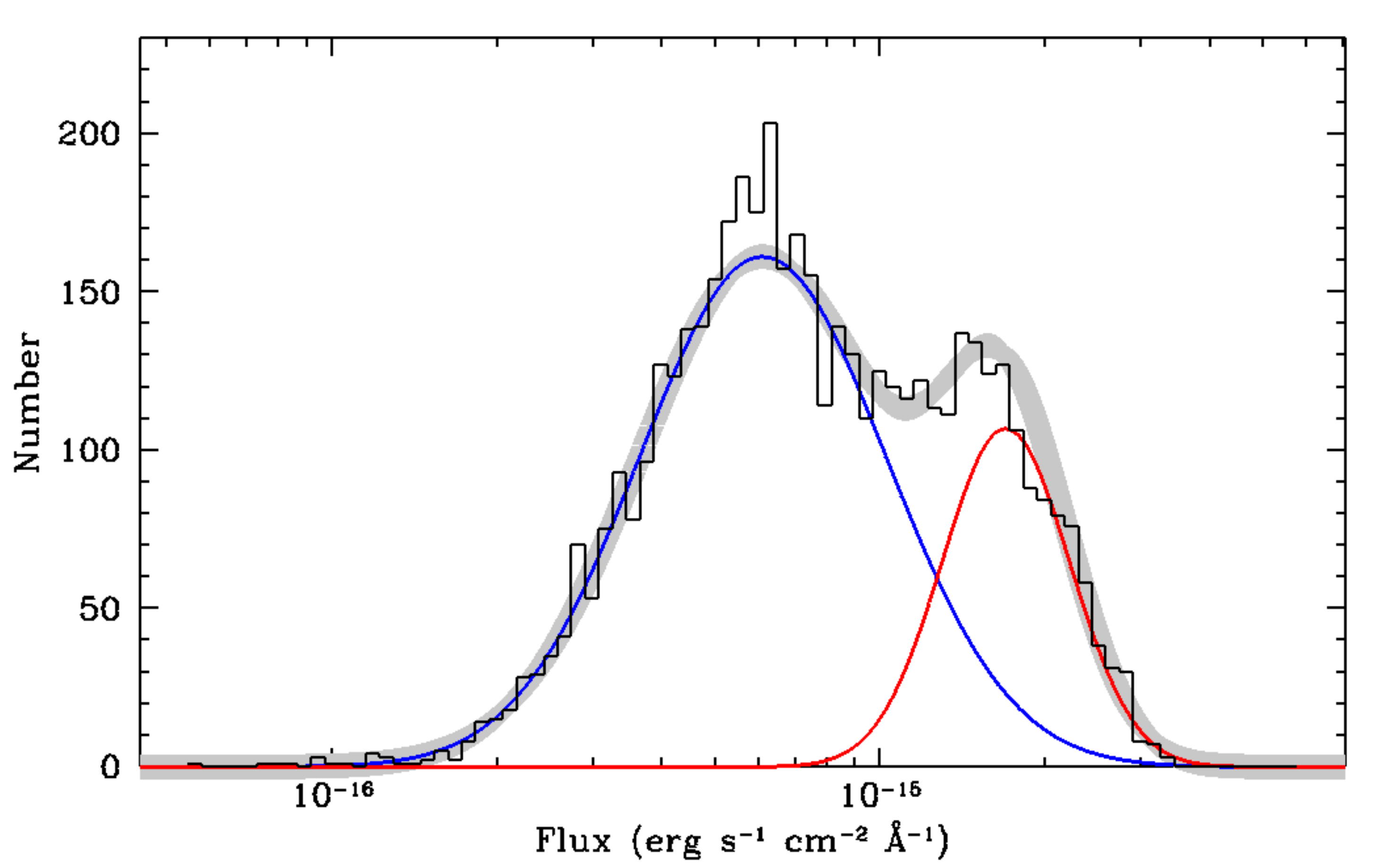}
\caption{{\bf The logarithmic far-UV flux distribution of J1858 during our observations.} The distribution is clearly bimodal, consistent with the visual impression from the light curve (Figure~\ref{fig:LCs} lower panel) of the variability being due to a flaring component that is superposed on a roughly constant component. The grey line is the optimal decomposition of the distribution into two Gaussians, as suggested by the KMM algorithm\cite{Bimodality_KMM1994AJ....108.2348A}. The blue and red lines correspond to the individual Gaussians. KMM rejects the null hypothesis of a single component with extremely high significance ($p < 10^{-43}$).}
\label{fig:lc_hist}
\end{figure}

\begin{figure}[ht]
\centering
\includegraphics[width=0.95\linewidth]{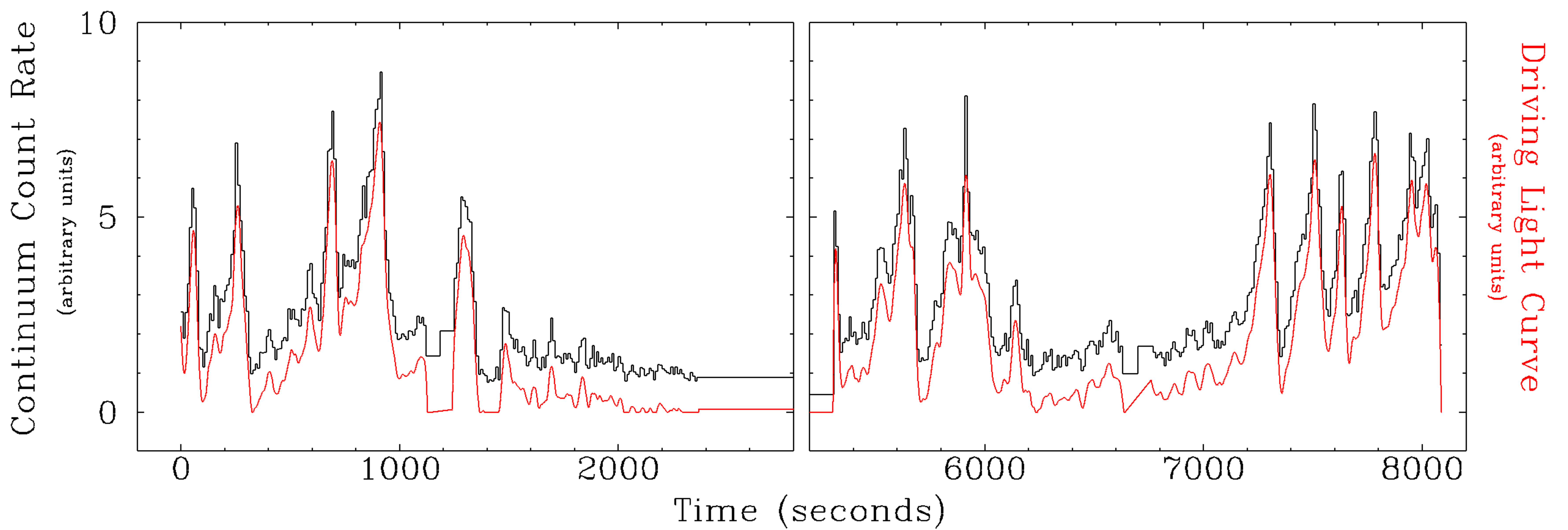}
\caption{{\bf The far-UV continuum and driving light curves.}
The black histogram shows the light curve of \src constructed from three broad wavelength regions that exclude the three strongest emission lines (N~{\sc v}~$\lambda1240$, Si~{\sc iv}~$\lambda1400$ and He~{\sc ii}~$\lambda1640$). The specific regions used were $\lambda\lambda$1290~{\AA} -- 1390~{\AA}, 1410~{\AA} -- 1630~{\AA}, 1660~{\AA} -- 1850~{\AA}. The light curve is shown normalized to an estimate of the underlying constant level (80 c~s$^{-1}$). The driving light curve used in the decomposition, $D(t)$, was constructed from this and is shown as the red curve. It was obtained by subtracting the estimate of the constant level, setting any slightly negative values to zero, and using a 5-point, 2nd order Savitzky-Golay filter to produce a slightly smoother, higher S/N version of the light curve.}
\label{fig:fdrive}
\end{figure}

\subsection*{Code availability}
Codes used for the analysis are available from the corresponding author upon reasonable request.

\subsection*{Data availability}
The data underlying this article are publicly available in: \url{https://archive.stsci.edu/hst/search.php} program ID 15984 for HST/FUV data.  \url{http://archive.eso.org/cms.html} program 190ID 2103.D-5052(A) for VLT/X-Shooter and \url{https://gtc. sdc.cab.inta-csic.es/gtc/}  program ID GTC23-19A for GTC/OSIRIS. X-ray data from NICER used all the OBSIDs starting with 120040, 220040,320040 and 359201 accessible from HIESARC (\url{https://heasarc.gsfc.nasa.gov/docs/nicer/nicer_archive.html}).

\end{document}